\documentclass[useAMS,usenatbib]{mn2e}
\usepackage{amsmath,amssymb}
\usepackage{natbib}
\usepackage{graphicx}
\usepackage{times}

\def\ifnote{\iftrue}

\title[GRB 080319B]{Clues from the prompt emission of GRB 080319B}
\author[Y. C. Zou, T. Piran \& R. Sari]{Yuan-Chuan Zou$^{1,2}$, Tsvi Piran$^{1}$ and Re'em Sari$^{1}$ \thanks{Email: zouyc@hust.edu.cn (YCZ), tsvi@phys.huji.ac.il (TP) and sari@tapir.caltech.edu (RS)}
\\
$^{1}${The Racah Institute of Physics, Hebrew University, Jerusalem 91904, Israel}\\
$^{2}${School of Physics, Huazhong University of Science and Technology, 430074 Wuhan, China}
}

\begin{document}
\date{\today}
\maketitle
\label{firstpage}

\begin{abstract}
The extremely bright optical flash that accompanied GRB 080319B suggested, at first glance, that  the prompt $\gamma$-rays in this burst were  produced by Synchrotron self Compton (SSC).   We analyze here the observed optical and $\gamma$ spectrum. We find that the very strong optical emission poses, due to self absorption, very strong constraints on the emission processes  and put  the origin of the optical emission at a very large radius, almost inconsistent with internal shock. Alternatively it requires a very large random Lorentz factor for the electrons. We find that  SSC could not have produced the prompt $\gamma$-rays. We also show that the optical emission and the $\gamma$ rays could not have been produced by synchrotron emission from two populations of electron within the same emitting region. Thus we must conclude that the optical and the  $\gamma$-rays were produced in different physical regions.   A possible interpretation of the observations is that the $\gamma$-rays arose from internal shocks but the optical flash  resulted from external shock emission. This would have been consistent with the few seconds  delay observed between the optical and $\gamma$-rays signals. 

\end{abstract}

\section{Introduction}
The radiation mechanism for the $\gamma$-ray burst (GRB) afterglow is widely accepted as synchrotron emission \citep[see][for a review]{P04}. However, the mechanism for the prompt emission is still uncertain. Most recently, \citet{km08} showed the inconsistency with the overall synchrotron model. On the other hand \citet{psz08} found that  Synchrotron self-Compton (SSC) cannot explain the prompt emission unless the prompt optical emission  is  very high. 
With a naked eye (5th magnitude) optical flash \citep{Cwiok08, Karpov08a}  GRB 080319B \citep{Racusin08b}, was a natural candidate for SSC \citep{kp08}.  

GRB 080319B  was located at redshift $z=0.937$ \citep{Vreeswijk08}. Its duration
$T_{90}$ was $\sim 57$s. The peak flux is $F_p \sim 2.26 \pm 0.21 \times 10^{-5} {\rm erg \,cm^{-2}s^{-1}}$ and the peak of the $\nu F_{\nu}$
spectrum $E_p \simeq 675{\pm 22}$keV (i.e., $\nu_p \sim 1.6\times 10^{20}$Hz, and consequently $F_{\nu,p} \sim 1.4 \times 10^{-25} {\rm erg\, cm^{-2} Hz^{-1} s^{-1}}$).  The photon indexes below and above  $E_p$ are $-0.855^{+0.014}_{-0.013}$ and
$-3.59^{+0.32}_{-0.62}$ respectively \citep{r08}. Choosing standard
cosmological parameters $H_0=70 {\rm km}\,{\rm s}^{-1}\,{\rm
Mpc}^{-1}, \Omega_m=0.3, \Omega_\lambda=0.7$, GRB 080319B had  a peak
luminosity $L_p \sim 9.67\times 10^{52} {\rm erg}\,{s}^{-1}$ and an
isotropic equivalent energy $E_{\rm \gamma, iso} \simeq 1.32 \times 10^{54}$
erg \citep{ga08}.

The optical observations were going on even before the onset of the
GRB because TORTORA was monitoring the same region of
sky at that moment. \citet{Karpov08a} reported the optical V-band
light curve in the prompt phase (from $\sim$ -10 s to $\sim$ 100 s).
Variability was evident and there were at least 3 or 4 pulses in
the optical light curve. The peak V-band magnitude reached 5.3,
corresponding to a flux density $\sim 28.7$ Jy.

Models for the optical emission accompanying the prompt $\gamma$-rays have been
extensively discussed since the early work of \citet{Katz94} who used  a low energy spectrum
$F_\nu \propto \nu^{1/3}$ and found that the prompt optical emission
could be bright to 18th mag. 
The observations of a prompt strong optical flash from GRB 990123 \cite{Akerlof99} lead to a 
wave of interest in this phenomenon.  The $t^{-2}$ decline of the early optical flash favored
a reverse shock model \citep{sp99,mr99} that was suggested just few month earlier
\citep{SP99a}. Optical emission from later internal shocks and residual internal shocks 
 have been discussed by \citet{wyf06} and \citet{lw08} respectively, 
 and the late internal shocks model has been
used to interpret the optical flares detected in GRB 990123, GRB
041219a, GRB 050904 and GRB 060111B
\citep{Akerlof99,Blake05,Boer06,Klotz06,zdx06}. However, none of the
previous bursts had the data as plentiful as that of GRB 080319B and
the constraints on the models were not very tight.

The very strong optical flash of  GRB 080319B lead naturally to the suggestion \citet{kp08,r08} that  internal shocks  synchrotron  produced the prompt 
optical emission while  SSC  produced the prompt  $\gamma$-rays.  We show here
that self absorption of the optical poses major constraints on the source of the optical emission and we consider its implications on general models (including SSC) for the emission of this burst. 
The paper is
structured as follows: In section 2 we consider the general constraints that arise from the optical emission. In section 4 we consider SSC model and in section 5  we make general remarks on
Inverse Compton (IC).  In section 6 we examine the possibility that the optical and $\gamma$-rays arose from two synchrotron emitting populations of electrons but in the same physical region. 
We find that the optical and the $\gamma$-rays were originated in physically different regimes. 
We discuss the implications of these results in section 6.

\section{Optical emission} \label{sec:opt}

The very strong optical flash that accompanied GRB 080319B poses the strongest constraints on the emission mechanism. A lot can be learnt from studying this flash on its own.
The observed optical signal, $F_{\nu, \rm{opt}}$,  must be less or equal than the corresponding black body 
emission:
\begin{equation}
F_{\nu, \rm{opt}} \le F_{BB} = 
    2 (1+z)^3 \nu^2_{\rm{opt}} \Gamma \gamma_e m_e ( \frac{R}{\Gamma d_L})^2 = 
{1.1 \times 10^{-24} \gamma_{e,2} R_{15}^2} {\Gamma_3}^{-1},   
    \label{BB}
    \end{equation}
where $m_e$ and $q_e$ (that we use later) are the electron's rest mass and charge, $\Gamma$ is the bulk Lorentz factor,  $\gamma_e$ is the typical electron Lorentz factor
$(\gamma_e m_e c^2 \approx kT)$, R is the emission radius, and $d_L$ is the luminosity distance\footnote{We denote by $A_i$ the quantity $A/10^i$ in cgs units.}.  This value should be compared with the observed optical flux $F_{\nu, \rm{opt}}  \sim 2.9 \times 10^{-22} {\rm erg\,cm^{-2}\,Hz^{-1}\,s^{-1}}$ which is more than  two orders of magnitude larger than the one found for $F_{BB}$ with ``typical" values. This is the essense of the problem of finding a reasonable solution for the emission mechanism in GRB080319B. 
By itself this constraint imposes a rather large $\gamma_e$ for reasonable values of $R$ and $\Gamma$, or alternatively a very large value of $R$. It will be the major constraint over which models that we examine later fail. 

This equation can be combined now with two expression that link $R$ and $\Gamma$:
The angular time scale:
\begin{equation}
\delta t >  {R \over 2 \Gamma^2 c},
\label{dt}
\end{equation}
and the deceleration radius:

\begin{equation}
R < R_{\gamma} = 
\left\{
\begin{array}{ll}
   \frac{E}{ 4 \pi A \Gamma^2 m_p c^2 }, &  {\rm wind},  \\
   \left(\frac{3E}{4 \pi n \Gamma^2 m_p c^2}\right)^{1/3}, &  {\rm ISM},
\end{array}
\right. \label{rgamma}
\end{equation}
where $m_p$ is the proton's mass, $E$ is the energy of the outflow, $n$ is the ISM density and $A$ is the wind parameter. 

\begin{figure}
 \includegraphics[width=0.5\textwidth]{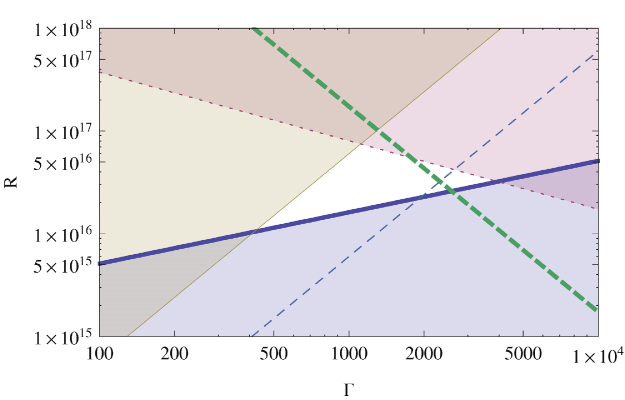}
 \caption{The parameter space in the  ($\Gamma,R$) plane that satisfies the  constraints Eqs. (\ref{BB}-\ref{rgamma}).  The thick solid line is  the black body constraint  for $\gamma_e=100$. The thin solid line indicates the variability time scale (for $\delta t_{opt}=$1 sec). Also shown (thin dashed line) is the variability constraint for $\delta t = 0.1$sec corresponding to the $\gamma$-ray variability. The thin dotted line and the thick dashed line are  the deceleration radius constraints  for ISM   and wind environments respectively. We choose conservatively  $ E=10^{55} {\rm ergs}$, $n=1 {\rm cm^{-3}}$,  and  $A=3\times 10^{33} {\rm cm}^{-1}$. The allowed regions for ISM with $\gamma_e=100$ is the white triangle at the center.  It increases slightly for a wind environment.  }
 \label{Fig:opt}
\end{figure}

Fig. \ref{Fig:opt} depicts the allowed region in the ($\Gamma,R$) phase space that satisfies all three constraints for either wind or ISM environments.
One can see that  the Black Body limit Eq. (\ref{BB}) pushes the emitting radius to large values.
On the other hand the two other constraints limits R to small values. The allowed region is  rather small and the radii are typically large and they won't be consistent with those needed for emitting the  $\gamma$-rays. Note that the allowed region  shrinks to zero if we take $\delta t \le 0.1$sec  as implied from the $\gamma$-ray observations. 

As the condition $ R <  R_\gamma$ applies for internal shocks, this result on its own limits strongly the ability of internal shocks to produce this optical flash. The only way out, within 
internal shocks is to increase $\gamma_e$ to very large values. However, typical internal shocks involve modest relativistic collisions in which such high Lorentz factors are not common (see however \cite{ks01}).

\section{Synchrotron Self-Compton}\label{sec:SSC}

We consider an SSC model with minimal assumptions. In fact unlike the previous section we don't use the inequalities (\ref{dt}) and (\ref{rgamma}) which depends on the overall model and we consider only the conditions within the emitting regions.  The low energy (including optical) emission is produced by synchrotron and the $\gamma$-rays are the inverse Compton of this synchrotron emission by the same electrons.  We assume that the emitting region is homogenous and  it moves radially outwards with a relativistic Lorentz factor $\Gamma$ towards us. It contains $N_e$ electrons with a typical Lorentz factor $\gamma_e$. Sightly generalizing we allow for  
a filling factor $f$  which implies that   only a fraction $f$ of the electrons
are emitting synchrotron while all the electrons involved in IC. This can happen, for example if the magnetic field $B$ occupies only a fraction $f$ of the volume. As we see later this helps but does not yield a satisfactory solution. 

\begin{figure}
 \includegraphics[width=0.5\textwidth]{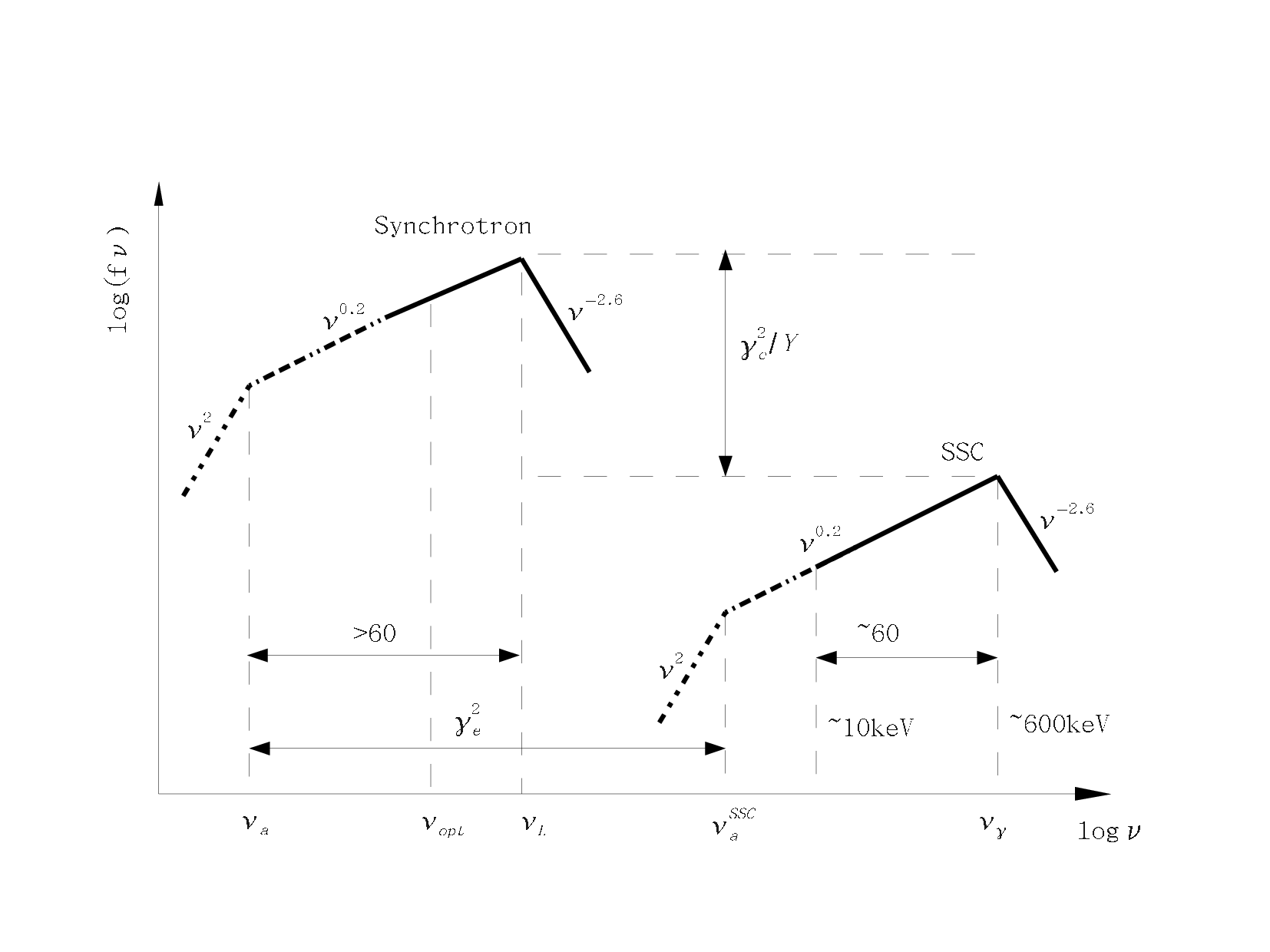}
 \caption{A schematic description of the spectrum in an SSC model.  Note that if 
$10 {\rm keV} /\gamma_e^2 > \nu_{opt}$,  that is if  $\gamma_e$ is small enough,  $\nu_{opt}$ might be below $\nu_{a}$ and in this case Eq. (\ref{alpha}) should be modified. }
 \label{Fig:schematic}
\end{figure}

We have four observables:  $ F_{\nu,\gamma}$, $F_{\nu, \rm{opt}}$,
$\nu_{\gamma}$, $\nu_{\rm{opt}}$  the fluxes and frequencies at the $\gamma$ and the optical.  We explore what are the conditions $(N_e, B, \Gamma, \gamma_e$ and $R$) needed to produce  the observations (see \cite{psz08} for a related approach).  As we assume that the $\gamma$-rays are produced by IC we have
\begin{equation}
Y \nu_L F_{\nu,L} = \nu_\gamma F_{\nu,\gamma}
\label{Y}
\end{equation}
and
\begin{equation}
\nu_\gamma = \gamma_e^2 \nu_L , 
\label{nu}
\end{equation}
where $F_{\nu,L} $ and $\nu_L$ are the (unknown) peak flux and peak frequency of the low energy component (that is being Inverse Compton scattered to produce the soft $\gamma$), $Y = \gamma_e^2 \tau$ is the Compton parameter and $\tau$ is the optical depth for Thompson  scattering. 
As the spectral shape is preserved by IC we can use the 
the spectral indices below and above the peak $\gamma$-ray frequency
 $\sim 0.2$ and $\sim -2.6$, respectively, to relate the optical flux and the flux at  $\nu_L$ as\footnote{Assuming that $\nu_L$ and $\nu_{opt}$ are in the same spectral regime. See subsequent discussion.} :
\begin{equation}
F_{\nu,opt} = F_{\nu,L} \left(\frac{\nu_{opt}}{\nu_L}\right)^\alpha ,
\label{alpha}
\end{equation}
where $\alpha$ is the observed index at the soft $\gamma$ range. 
$\alpha$ will be either $0.2$ or $-2.6$ depending on whether $\nu_L$ is larger
or smaller than $\nu_{opt}$ (which we call UV and IR solutions respectively). 
The overall spectral distribution is shown in Fig. \ref{Fig:schematic}.

Combining these three equations we get:
\begin{equation}
\gamma_e = \left(\frac{Y F_{\nu,opt}~\nu_\gamma^\alpha}{F_{\nu,\gamma} ~\nu_{opt}^\alpha}\right)^{\frac{1}{2(1+\alpha)}}
= 530 \left(\frac{Y}{140}\right)^{\frac{1}{2(1+\alpha)}} .
\label{eq:gamma_e-Y}
\end{equation}
We have $\gamma_e  \sim 70 Y^{0.4}$ for $\alpha = 0.2$ corresponsing to a UV solution ($\nu_L > \nu_{opt}$ ) while for the 
 IR solution ($\nu_L < \nu_{opt}$ ) $\alpha = -2.6$ and   $\gamma_e \sim 2500Y^{-0.3}$.

Using this expression for  $\gamma_e$ 
we solve now the synchrotron frequency and peak synchrotron flux equations:
\begin{equation}
\nu_L =  \frac{1}{(1+z)} \frac{3 q_e}{2 \pi m_e c} \Gamma
B\gamma_e^2 
\end{equation}
and 
\begin{equation}
F_{\nu,L} =  \frac{m_e c^2 \sigma_T}{3 q_e} \frac{(1+z)}{4 \pi
d^2_L}   \Gamma B f N_e.
\end{equation}
Note that  $\Gamma$ and $B$ appears as a product and hence we have two  equations for 
two  variables  $N_e$ and $\Gamma B$  within the emitting region. Once we know $N_e$ we can solve for $R$, using the optical depth $\tau = Y/\gamma_e^2$ and remembering $\tau=\sigma_T N_e/4\pi R^2$:
\begin{equation}
 R = \frac{1}{1+z}\sqrt{\frac{3 d_L^2 q_e^2 F_{\nu,\gamma} \gamma_e^8}{2 \pi c^3 f m_e^2 {\nu_\gamma} Y^2}}  \approx 2 \times 10^{15} \frac{\gamma_{e,2}^4}{\sqrt{f} Y} .
\label{eq:R-Y}
\end{equation}
Turning now to the flux limit equation  (\ref{BB})  we find
\begin{equation}
{F_{BB} \over F_{\nu,opt}}=  \frac{(1+z) 3 q_e^2 {F_{\nu,\gamma}} \gamma_e^9 \nu_{opt}^2}{c^3 f {F_{\nu,opt}} \Gamma m_e {\nu_\gamma}  Y^2} =  0.06  \frac{  \gamma_{e,2}^9}{f \Gamma_3 Y^2} .
\end{equation}
Substitution of the   expression for $\gamma_e$ in this equation yields for $\alpha=0.2$:
\begin{equation} 
{F_{BB} \over F_{\nu,opt}}=2 \times 10^{-3} \frac{Y^{1.75}}{f \Gamma_3}.
\end{equation}
This ratio is  not sensitive to the value of $\alpha$ used. It is clear that for reasonable values of 
$\Gamma$ this ratio is less than unity unless $Y$ is extremely large.  This is the essence
of the optical self-absorption problem that forbids any low $Y$ SSC solution.  A large $Y$ will lead to a  energy crisis where most of the energy of this (already very powerful) burst would have been emitted in the GeV regime leading to a huge overall energy requirement. Note that the isotropic equivalent $\gamma$-ray energy of this bursts is already larger than $10^{54}$ergs.  

Recalling that there is no obvious break in the  $\gamma$-rays spectrum from the peak frequency $\sim 600$keV down to $\sim 10$keV, a self absorption break should not appear in the range [$\nu_L/60,\nu_L$]. This sets an even more powerful constraint:
$F_{BB}(\nu_L/60)> F_{\nu_L/60}$ which is typically significantly more difficult to satisfy (by ratio $(\nu_L/60\nu_{opt})^{2-\alpha}$ than Eq. (\ref{BB}). 

We consider now briefly three caveats for the above result. 

\noindent (i) The filling factor $f$ allows for a more reasonable solution. But one needs an extremely small $f$ for a valid one.  

\noindent (ii) In a very small region of the parameter phase space, that is if $10 {\rm keV}/\gamma_e^2 > \nu_{opt}$ (see Fig. \ref{Fig:schematic} )   an ``optically thick" UV solution is possible. In this case 
$F_{\nu}$ increases like $\nu^2$  from $\nu_{opt}$ to $\nu_L/60$ . 
The  solution is slightly different from the one given above (as Eq. (\ref{alpha}) has to be modified) but qualitatively the results remains unchanged (see Fig. \ref{Fig:schematic}). 

\noindent (iii) For the IR solution ($\alpha=-2.6$), from Eq. (\ref{eq:gamma_e-Y}) we find  $\gamma_e \sim 3 \times 10^3$. This leads to a huge emission  radius (see Eq.  (\ref{eq:R-Y})) which is larger than  $\sim 10^{20}$ cm and this solution can be easily ruled out.

\section{General Inverse Compton Model}\label{sec:IC}

So far we have considered an SSC solution. However, it is interesting to note that Eqs. (\ref{Y}-\ref{eq:gamma_e-Y}) apply when the prompt $\gamma$-rays are the IC scattering of low energy external  photons, provided that there is no significant relativistic bulk motion between the source of the seed photons and the IC electrons.  For the UV solution, $\gamma_e \sim 50$ implies that the total kinetic energy should be much larger than the energy of the prompt $\gamma$-rays as the total internal energy of the electrons is much less than the rest mass energy of protons. This leads to a severe  energy budget problem. For the IR solution ($\alpha = -2.6$)  the Lorentz factor of the electrons is  $\gamma_e \sim 2.5 \times 10^3 Y^{-0.31}$, then the peak of the low energy  emission is at $\nu_L =  2.5 \times 10^{13} Y^{0.625}$ and the needed peak flux density (not necessarily synchrotron emission) is $F_{\nu_L}  = 8.6 \times 10^{-19} Y^{-1.625} {\rm erg\, cm^{-2} Hz^{-1} s^{-1}}$. Given the very small $\nu_L$ one can hardly  satisfy $F_{\nu_L} \leq F_{BB}(\nu_L)$ (or even stronger  $F_{\nu_L/60} \leq F_{BB}(\nu_L/60)$) for reasonable parameters.

\section{Two Electron Populations}\label{sec:two synchrotrons}
We consider now an alternative model in which  the optical and the soft $\gamma$-rays arise from synchrotron emission from the same physical region but from different populations of electrons. We denote these populations  with subscripts L and $\gamma$ for the lower energy band and higher one respectively.  $\gamma$ and $N$ are the typical Lorentz factor and the total number of electrons respectively of each kind. 
As we assume a single emitting region  the magnetic field $B$ and the bulk Lorentz factor $\Gamma$ should be the same. 

Using the peak synchrotron frequency relation $\nu_p \propto \gamma^2 \Gamma B$
and the peak flux density $F_{\nu,p} \propto N \Gamma B$:
\begin{equation}
\frac{\gamma_L^2 }{\gamma_\gamma^2 } = \frac{\nu_L}{\nu_\gamma},
\label{eq:nup}
\end{equation}
and 
\begin{equation}
\frac{N_L }{N_\gamma } = \frac{F_{\nu,L}}{F_{\nu,\gamma}}
\label{eq:fnup}
\end{equation}
The total isotropic internal energy should be $\propto N_e \gamma$. Combining the above two equations, we obtain:
\begin{equation}
\frac{\gamma_L N_L}{\gamma_\gamma N_\gamma} = \frac{\nu_L^{1/2} F_{\nu,L}}{\nu_\gamma^{1/2}F_{\nu,\gamma}} = \frac{\nu_{opt}^{1/2} F_{\nu,opt}}{\nu_\gamma^{1/2}F_{\nu,\gamma}} \left( \frac{\nu_L}{\nu_{opt}}\right)^{1/2+\tilde \alpha},
\label{eq:energy}
\end{equation}
where $\tilde \alpha$ is the spectral slope in the range [$\nu_{opt}, \nu_L$].
If $\nu_L>\nu_{opt}$, then clearly $(\nu_L/\nu_{opt})^{1/2+\tilde \alpha}  > 1$. On the other hand, if $\nu_L < \nu_{opt}$, $\tilde \alpha <-1$  and again $(\nu_L/\nu_{opt})^{1/2+\tilde \alpha}  > 1$.
Using the observed values we obtain:
\begin{equation}
\frac{\gamma_L N_L}{\gamma_\gamma N_\gamma}  \ge \frac{\nu_{opt}^{1/2} F_{\nu,opt}}{\nu_H^{1/2}F_{\nu,\gamma}} \sim 4 .
\label{eq:energy1}
\end{equation}
Thus a peculiar condition of this model is that the energy of the lower energy electron population that is responsible for producing the (relatively weak) optical signal  exceeds that of the $\gamma$-rays producing component.  Once more we are faced with an  energy budget problem that makes  the two synchrotron components model quite unlikely  (see also \citet{M08}).

\section{Conclusions and discussions}
\citet{psz08} have shown recently that typical GRBs  with normal (weaker than 12th magnitude) prompt  optical emission cannot be produced by SSC of a softer  component.  The unique burst GRB 080319B had a very luminous prompt optical emission and one could expect, at first glance that the prompt $\gamma$-rays were produced in SSC of the optical prompt emission. This was the accepted interpretation in the discovery paper \citep{r08} as well as in several others \citep{kp08,fp08}.   The numerous detailed  observations of this burst led to the hope that one can determine the physical parameters within the emitting regions here. However, a  careful analysis reveals a drastically different picture. There is no reasonable SSC solution. One can make an even stronger statement and argue that it is unlikely that the prompt $\gamma$-rays are produced by inverse Compton scattering of seed photons produced by a source with no relativistic bulk motion relative to the IC electrons (regardless of the origin of the seed photons). 

We also considered a situation with two populations of electrons that co-exist in the same emitting region. Both populations emit synchrotron radiation with the less energetic electrons producing the optical while the more energetic ones produce the $\gamma$-rays. We have shown that even though the total energy released in the optical is orders of magnitude lower than the energy released in $\gamma$-rays the lower energy electrons population should carry more energy - making, once more, the model energetically inefficient.  

Combining these two results we conclude that   the optical emission and the soft $\gamma$-rays do  not come from a single origin. This is the main conclusion of this letter. Once we relax the condition that both modes of prompt emission are produced in the same region,
there are  many possibilities. However, even here the  limits obtained in section \ref{sec:opt} indicate that the optical emission is produced at a very large radius which is most likely incompatible with internal shocks (Note that typical internal shocks will have a rather modest values of $\gamma_e$.). This raises the possibility that the optical emission is produced in this
burst by the early external shock (possibly by the reverse shock). While there are several problems with this model (in particular the fast rise of the optical emission that is faster than expected in this case \cite{np04}) they are less severe than those  basic issues that arise with the
internal shocks.  The fact that the optical emission follows to some extend the $\gamma$-rays but with a time delay of several seconds \citep{Guidorzi08} is consistent with this model.

\section*{Acknowledgments}
We thank P. Kumar, R. Mochkovitch and Y. Fan for discussion. This work is supported by the Israel Science Foundation center of excellence in High energy astrophysics and by the Schwartzmann University Chair (TP), by A Marie Curie IRG and NASA ATP grant (RS), by the National Natural Science Foundation of China grant 10703002 and  a  Lady Davis Trust post-doctorical fellowship (YCZ).

\end{document}